\documentclass[a4paper,11pt]{article}
\usepackage{latexsym}
\usepackage{epsfig}
\usepackage{amssymb}
\author{H. Mohseni Sadjadi \footnote{mohsenisad@ut.ac.ir} and  Parviz Goodarzi  \footnote{p\underline\space goodarzi@ut.ac.ir}
\\ {\small Department of Physics, University of Tehran,}
\\ {\small P. O. B. 14395-547, Tehran 14399-55961, Iran}}
\title{Temperature in warm inflation in non minimal kinetic coupling model}

\begin{document}

\maketitle
\begin{abstract}
Warm inflation in the non minimal derivative coupling model with a general dissipative coefficient is considered.
We investigate conditions for the existence of the slow roll approximation and study cosmological perturbations. The spectral index, and the power spectrum are calculated  and the temperature of the universe at the end of the slow roll warm inflation is obtained.
\end{abstract}

\section{Introduction}

To describe the inflationary phase in the early universe \cite{guth,inflaton1}, many theories have been proposed which most of them are categorized into two classes: modified gravity models \cite{stari}, and models with exotic fields dubbed as inflaton \cite{exo}. These groups may related to each other through some conformal transformations \cite{conf}.

In a well known model, the responsible of the early accelerated expansion of the Universe is a canonical scalar field $\varphi$, rolling down slowly a nearly flat potential. Inflation lasts as long as the slow roll conditions hold. In this paradigm we encounter a cold universe at the end of inflation. After the cease of the slow roll conditions, the scalar field begins a rapid coherent oscillation and decays to ultra relativistic particles (radiation) reheating the Universe \cite{reh}. A natural candidate for this scalar field, as is proposed in \cite{berz}, is the  Higgs boson. In this context, adding a non-minimal coupling between the scalar field and scalar curvature is required for the renormalizability,
and also consistency with the amplitude of density perturbations obtained via observations.
Another model in which the inflaton is considered as the Higgs field is introduced in \cite{Germani1}, where the scalar field has a non minimal kinetic coupling term.  This theory does not suffer from unitary violation and is safe of quantum corrections.  In this framework, the inflation and the reheating of the Universe are discussed in the literature \cite{sadreh}. The same model, with a non canonical scalar field dark energy,  is also employed to describe the present acceleration of the Universe \cite{sad3}. In the aforementioned model, inflation and reheating happen in two distinct eras, but one can unify them by assuming an appropriate dissipative coefficient which permits the decay of inflaton to radiation during inflation:
Warm inflation was first introduced for minimal coupling model \cite{Berera}. Afterwards, numerous articles has been published in this subject \cite{Herrera,Cai,Hosoya}. Friction term for inflaton equation of motion is computed in \cite{Hosoya}. Tachyonic warm inflationary universe models are considered in \cite{Herrera2}.

In this work we consider warm inflation in non minimal derivative coupling model.  In the second and third sections, based on our previous papers \cite{sadreh}, we review the nonminimal derivative coupling model in the presence of an additional radiation sector and investigate slow roll conditions. In the fourth section, the perturbations in the model are studied, and the discussion is conducted in such a way that the parameter extractable from the observations such as the spectral index acquire more utilizable and more general compact form with respect to \cite{noz}, where the perturbations of this model were also discussed.  By employing PLANK2013 data,  we use our results to obtain the temperature at the end of warm slow roll inflation.

 We use units $\hbar=c=8\pi G=1$ though the paper.

\section{Preliminaries}

The action of Gravitational Enhanced Friction (GEF) theory is given by \cite{Germani1}
\begin{equation}\label{1}
S=\int \Big({1\over 2
}R-{1\over 2}\Delta^{\mu \nu}\partial_\mu
\varphi \partial_{\nu} \varphi- V(\varphi)\Big) \sqrt{-g}d^4x+S_{int}+S_{r},
\end{equation}
where $\Delta^{\mu \nu}=g^{\mu \nu}+{1\over M^2}G^{\mu \nu}$, $G^{\mu \nu}=R^{\mu \nu}-{1\over 2}Rg^{\mu \nu}$ is Einstein
tensor, $M$ is a constant, $S_{r}$ is the matter action and $S_{int}$ describes the interaction of the scalar field with all other ingredients. In the absence of terms containing more than two time derivatives, we have not additional degrees of freedom in this theory. We calculate the energy momentum tensor,
\begin{equation}\label{1.1}
T_{\mu\nu}=T^{(\varphi)}_{\mu\nu}+T^{(r)}_{\mu\nu},
\end{equation}
by variation of the action with respect to the metric \cite{late}. $T^{(r)}_{\mu\nu}$ is the radiation energy momentum tensor and $T^{(\varphi)}_{\mu\nu}$ is the scalar field energy momentum tensor, consisting of parts coming from the minimal part: $\mathcal{T}_{\mu\nu}$,
\begin{equation}\label{1.2}
\mathcal{T}^{(\varphi)}_{\mu\nu}=\nabla_{\mu}\varphi\nabla_{\nu}\varphi-{1\over2}g_{\mu\nu}{(\nabla\varphi)}^2-g_{\mu\nu}V(\varphi),
\end{equation}
and parts coming from the non minimal derivative coupling section, $\Theta_{\mu\nu}$,
\begin{eqnarray}\label{1.3}
&&\Theta_{\mu\nu}=-{1\over2}G_{\mu\nu}{(\nabla\varphi)}^2-{1\over2}R\nabla_{\mu}\varphi\nabla_{\nu}\varphi
+R^{\alpha}_{\mu}\nabla_{\alpha}\varphi\nabla_{\nu}\varphi\\\nonumber
&&+R^{\alpha}_{\nu}\nabla_{\alpha}\varphi\nabla_{\mu}\varphi+R_{\mu\alpha\nu\beta}\nabla^{\alpha}\varphi\nabla^{\beta}\varphi
+\nabla_{\mu}\nabla^{\alpha}\varphi\nabla_{\nu}\nabla_{\alpha}\varphi\\\nonumber
&&-\nabla_{\mu}\nabla^{\nu}\varphi\Box \varphi
-{1\over2}g_{\mu\nu}\nabla^{\alpha}\nabla^{\beta}\varphi\nabla_{\alpha}\nabla_{\beta}\varphi+{1\over2}g_{\mu\nu}{(\Box\varphi)}^2\\\nonumber
&&-g_{\mu\nu}\nabla_{\alpha}\varphi\nabla_{\beta}\varphi R^{\alpha\beta}.
\end{eqnarray}

By variation of the action (\ref{1}) with respect to the scalar field $\varphi$,  the equation of motion for the homogeneous and isotopic scalar field in the presence of
a dissipative term can be expressed as
\begin{equation}\label{2}
(1+{3H^2\over M^2})\ddot{\varphi}+3H(1+{3H^2\over M^2}+{2\dot{H}\over M^2})\dot{\varphi}+V'(\varphi)+\Gamma\dot{\varphi}=0,
\end{equation}
where $H={\dot{a}\over a}$ is the Hubble parameter, a "dot"  is the differentiation with respect to the cosmic time $t$, "prime" is differentiation with respect to the scalar field $\varphi$,  and $\Gamma\dot{\varphi}$ is the friction term adopted phenomenologically to describe decay of the $\varphi$ field and its energy transfer into the radiation bath.  $\Gamma$ in general is a function of $\varphi$ and temperature \cite{Berera2,Xiao}. The Friedman equation for this model is given by
\begin{equation}\label{3}
H^2={1\over 3}((1+{9H^2\over M^2}){\dot{\varphi}^2\over 2}+V(\varphi)+\rho_{r}),
\end{equation}
where $\rho_{r}$ is the energy density of the radiation, which can be written as \cite{Berera2}
\begin{equation}\label{4}
\rho_{r}={3\over4}TS.
\end{equation}
$S$ is the entropy density and $T$ is the temperature. The energy density and pressure of homogenous and isotropic scalar field are given by
\begin{equation}\label{4.1}
\rho_{\varphi}=((1+{9H^2\over M^2}){\dot{\varphi}^2\over 2}+V(\varphi)),
\end{equation}
and
\begin{equation}\label{4.2}
P_{\varphi}=(1-{3H^2\over M^2}-{2\dot{H}\over M^2}){\dot{\varphi}^2\over 2}-V(\varphi)-{2H\dot{\varphi}\ddot{\varphi}\over M^2},
\end{equation}
respectively.
By continuity equation for the total system $\dot{\rho}+3H(\rho+P)=0$, and also the equation of motion (\ref{2}),  we obtain
\begin{equation}\label{6}
\dot{\rho_{r}}+4H\rho_{r}=\Gamma{\dot{\phi}}^2,
\end{equation}
which gives the rate of entropy production as
\begin{equation}\label{5}
T(\dot{S}+3HS)=\Gamma\dot{\varphi}^2.
\end{equation}

\section{Slow roll approximation}

In the previous section we pointed out to the equations needed to describe the scalar field and radiation evolutions in an interacting nonminimal coupling model.  Hereafter we consider the slow roll approximation:
\begin{equation}\label{7}
\ddot{\varphi}\ll3H\dot{\varphi} \qquad  \dot{H}\ll H^2 \qquad (1+{9H^2\over M^2}){\dot{\varphi}^2\over 2}\ll V(\varphi).
\end{equation}
The entropy density satisfies
\begin{equation}\label{7.1}
TS\ll V(\varphi) \qquad \dot{S}\ll 3HS.
\end{equation}

For a positive potential, the slow roll conditions give rise to the inflation. Neglecting the second order derivative, we can write the equation of motion of the scalar field as
\begin{equation}\label{8}
\dot{\varphi}\simeq -{V'(\varphi)\over{3HU(1+r)}},
\end{equation}
where
\begin{equation}\label{8.1}
U=1+{3 H^2\over M^2} \qquad r={\Gamma\over 3UH}.
\end{equation}
$r$ is the ratio of thermal damping component to the expansion damping.
During the slow roll warm inflation, the potential energy of the scalar field is dominant, and therefore
the Friedman equation becomes
\begin{equation}\label{9}
H^2\simeq {1\over 3} V(\varphi).
\end{equation}
We have also
\begin{equation}\label{9.1}
ST\simeq Ur\dot{\varphi}^2.
\end{equation}
By equation (\ref{9}) we can write $U$ as a function of the potential
\begin{equation}\label{10}
U=1+{V(\varphi)\over M^2 }.
\end{equation}
We employ the following set of parameters to characterize the slow roll:
\begin{equation}\label{11}
\delta={1\over 2}{\left({V'(\varphi)\over V(\varphi)}\right)}^2{1\over U(\varphi)},
\end{equation}
\begin{equation}\label{12}
\eta={V''(\varphi)\over V(\varphi)}{1\over U(\varphi)},
\end{equation}
\begin{equation}\label{13}
\beta={\Gamma'(\varphi)V'(\varphi)\over\Gamma(\varphi)V(\varphi)}{1\over U(\varphi)},
\end{equation}
\begin{equation}\label{14}
\epsilon=-{\dot{H}\over H^2}.
\end{equation}
To express slow roll conditions in terms of these parameters, we need to calculate $\dot{U}$ and $\dot{r}$. We have
\begin{equation}\label{15}
\dot{U}={6\dot{H}H\over M^2},
\end{equation}
therefore
\begin{equation}\label{16}
{\dot{U}\over H}=-2\epsilon (U-1),
\end{equation}
and
\begin{equation}\label{16.1}
{\dot{r}\over H}=-\beta{r\over r+1}+\epsilon r(3-{2\over U}).
\end{equation}
Using the relation (\ref{9}), one can  obtain $\epsilon$ as a function $\delta$ and $r$
\begin{equation}\label{17}
\epsilon={\delta\over 1+r}.
\end{equation}
From (\ref{8}) we can derive
\begin{equation}\label{18}
{\ddot{\varphi}\over H\dot{\varphi}}=-\eta{1\over r+1}+\delta (3+{2\over U}){1\over (1+r)^2}+\beta{r\over(1+r)^2}.
\end{equation}
The slow roll conditions can be expressed as
\begin{equation}\label{22}
\epsilon \ll 1,\,\,\,  \delta \ll 1+r,\,\,\,  \eta \ll 1+r, \,\,\, \beta \ll 1+r.
\end{equation}
Note that if ${H^2\over M^2}\rightarrow 0$  our model reduces to warm inflation in minimal coupling model\cite{Berera}, and if $r\rightarrow 0$ and ${H^2\over M^2}\rightarrow 0$ we recover the standard slow roll inflation \cite{lid}.  By using the relations (\ref{11}), (\ref{12}), and (\ref{13}),  we get:
\begin{equation}\label{19}
{1\over H}{dln(TS)\over dt}=\epsilon(1+2{(3-{2\over U})\over1+r})+\beta{-1+r\over(1+r)^2}-2\eta{1\over(1+r)}.
\end{equation}
In our study, we take $r\gg1$ and consider the high friction limit
\begin{equation}\label{highfriction}
{H^2\over M^2}\gg 1,
\end{equation}
therefore $U\simeq {3H^2\over M^2}\gg1$ and ${1\over H}{dln(TS)\over dt}={1\over H}\left({\dot{T}\over T}+{\dot{S}\over S}\right)\ll 1$.

The number of efolds during slow roll warm inflation is
\begin{equation}\label{20}
\mathcal{N}=\int^{t_{end}}_{t_{\star}} Hdt=\int^{\varphi_{end}}_{\varphi_{\star}}{H\over\dot{\varphi}}d\varphi=
-\int^{\varphi_{end}}_{\varphi_{\star}}{3H^2U(1+r)\over V'(\varphi)}d\varphi,
\end{equation}
where $\varphi_{\star}=\varphi(t_{\star})$ and $\varphi_{end}=\varphi(t_{end})$ are the values of the scalar field at the horizon crossing ($t_{\star}$), and at the end of inflation, ($t_{end}$).  By horizon crossing (or horizon exit) we mean the time at which a pivot scale exited the Hubble radius during inflation.
Using the Friedman equation the above relation becomes
\begin{equation}\label{21}
\mathcal{N}=\int^{\varphi_{end}}_{\varphi_{\star}}{V(\varphi)\over V'(\varphi)}U(1+r) d\varphi.
\end{equation}
At the end of this section, by choosing the form of $\Gamma$ and the potential, we derive more specific results.
We adopt the (general) damping term  proposed in \cite{Berera2}
\begin{equation}\label{23}
\Gamma=\Gamma_0{({\varphi\over\varphi_0})}^p,
\end{equation}
where $p$ is an arbitrary real number and $\varphi_0,\Gamma_0$ are constant,
and consider the power law potential
\begin{equation}\label{24}
V(\varphi)=\lambda \varphi^n,
\end{equation}
where $n$ and $\lambda$ are two constants. By using relation (\ref{10}), and in high friction limit for $r\gg1$, after some computations we obtain
\begin{equation}\label{25}
\rho_r={\Gamma\dot{\varphi}^2\over 4H}.
\end{equation}
By inserting $\dot{\varphi}$ from (\ref{8}), into the above equation we obtain
\begin{equation}\label{26}
\rho_r={V'(\varphi)^2\over 4H\Gamma}={\sqrt{3} V'(\varphi)^2\over 4\Gamma\sqrt{V(\varphi)}}.
 \end{equation}
Using (\ref{23}) and (\ref{24}),  $\rho_r$ is obtained as
\begin{equation}\label{27}
\rho_r={\sqrt{3} n^2 \lambda^{3\over2}\varphi_0^p \over 4\Gamma_0}\varphi^{({3n\over2}-2-p)}.
\end{equation}

We can write radiation energy density as a function of temperature,
\begin{equation}\label{28}
\rho_r={g\pi^2\over 30}T^4,
\end{equation}
where $g$ is the number of degree of freedom for ultra relativistic particles.
By relations (\ref{27},\ref{28}) temperature of the universe may derived as a function of $\varphi$
\begin{equation}\label{29}
T=A \varphi^{{3n-4-2p\over 8}},
\end{equation}
where in this relation $A$ is given by
\begin{equation}\label{30}
A={\left({15\sqrt{3} n^2 \lambda^{3\over2}\varphi_0^p\over 2\Gamma_0 g\pi^2}\right)}^{1\over4}.
\end{equation}

The slow roll parameters may be now expressed as
\begin{equation}\label{31}
\delta={M^2 n^2\over 2\lambda}{1\over\varphi^{n+2}},
\end{equation}
\begin{equation}\label{32}
\eta={M^2 n(n-1)\over \lambda}{1\over\varphi^{n+2
}},
\end{equation}
and the number of efolds is given by
\begin{equation}\label{33}
\mathcal{N}={1\over 3}\int_{\varphi_{end}}^{\varphi_\star} {V(\varphi)\Gamma\over V'(\varphi)H}d\varphi,
\end{equation}
where $\varphi_{\star}=\varphi(t_{\star})$ and $t_{\star}$ is the time at the horizon crossing. By using (\ref{29}) and assuming $\varphi_{\star}\ll \varphi_{end}$, the number of e fold becomes
\begin{equation}\label{34.1}
\mathcal{N}={4\Gamma_0\over \sqrt{3}{n\sqrt{\lambda}\varphi_0^p}}\times{\varphi_\star^{4p-2n+8\over 4}\over 4p-2n+8}.
\end{equation}

${\ddot{a}\over a}=H^2(1-\epsilon)$ implies that the inflation ends when  $\epsilon\sim 1$ ($\sim$ denotes the order of magnitude).  Putting $\epsilon\sim 1$ back into (\ref{17}) gives $\delta\sim 1+r $ and if $r\gg 1 $,  at the end of warm inflation we have $\delta\sim r$.

\section{Cosmological perturbations}
In this section we consider the evolution equation for the first order cosmological perturbations of a system containing inflaton and radiation.
In the Newtonian gauge, scalar perturbations of the metric can be written as \cite{Weinberg}
\begin{equation}\label{35}
ds^2=-(1+2\Phi)dt^2+a^2(1-2\Psi)\delta_{ij}dx^idx^j.
\end{equation}
The energy momentum tensor splits into radiation  $T^{\mu\nu}_r$ and inflaton part $T^{\mu\nu}_{\varphi}$,
\begin{equation}\label{35.5}
T^{\mu\nu}=T^{\mu\nu}_r+T^{\mu\nu}_{\varphi}.
\end{equation}
$T^{\mu\nu}_{\varphi}$ is the energy momentum tensor of the inflaton, introduced in the second section.
We have modeled the radiation field as a  perfect barotropic fluid. We have
\begin{equation}\label{36}
T^{\mu\nu}_r=(\rho_{r}+P_r)u_{\mu}u_{\nu}+P_rg_{\mu\nu},
\end{equation}
where $u_r$ is four velocity of radiation fluid and $\overline{u}_i=0$ and $\overline{u}_0=-1$. "Bar" denotes unperturbed quantities. By considering the normalisation condition $g^{\mu\nu}u_{\mu}u_{\nu}=-1$, we obtain
\begin{equation}\label{37}
\delta u^0=\delta u_0={h_{00}\over 2}.
\end{equation}
$\delta u^i$  is an independent dynamical variable. We can define  $\delta u_i=\partial_i\delta u$ \cite{Weinberg}.
Energy transfer between the two components is described by a flux term \cite{Moss}
\begin{equation}\label{38}
 Q_{\mu}=-\Gamma u^{\nu}\partial_{\mu}\varphi \partial_{\nu}\varphi,
\end{equation}
associated to the field equations
 \begin{equation}\label{39}
\nabla_{\mu}T^{\mu\nu}_r=Q^{\nu},
\end{equation}
and
\begin{equation}\label{39.5}
\nabla_{\mu}T^{\mu\nu}_{\varphi}=-Q^{\nu}.
\end{equation}
From relation (\ref{38}) we deduce $Q_{0}=\Gamma \dot{\varphi}^2$, so the unperturbed equation (\ref{39}) becomes $Q_{0}=\dot\rho_r +3H(\rho_r+P_r)$ which is the continuity equation for radiation field in the presence of interaction.  Similarly, the equation (\ref{39.5}) becomes $-Q_{0}=\dot\rho_{\varphi} +3H(\rho_{\varphi}+P_{\varphi})$.
Perturbations to the energy momentum transfer are described by the energy transfer
\begin{equation}\label{40}
\delta Q_{0}=-\delta\Gamma\dot{\varphi}^2+\Phi\Gamma\dot{\varphi}^2-2\Gamma\dot{\varphi}\dot{\delta\varphi},
\end{equation}
and the momentum flux
\begin{equation}\label{41}
\delta Q_{i}=-\Gamma\dot{\varphi}\partial_i{\delta\varphi}.
\end{equation}
By variation of equation (\ref{39}) as $\delta(\nabla_{\mu}T^{\mu\nu}_r)=\delta Q^{\nu}$, for the zeroth (0-0) component we obtain
\begin{equation}\label{42}
\dot{\delta\rho_r}+4H\delta\rho_r+{4\over3}\rho_r\nabla^2\delta u-4\dot{\Psi}\rho_r=
-\Phi\Gamma{\dot\varphi}^2+\delta\Gamma{\dot\varphi}^2+2\Gamma\dot{\delta\varphi}\dot{\varphi},
\end{equation}
and for the $i-th$ component we derive
\begin{equation}\label{42.1}
4\rho_r\dot{\delta u^i}+4\dot{\rho_r}\delta u^i+20 H\rho_r\delta
u^i=-a^2[3\Gamma\dot{\varphi}\partial_i\delta\varphi+\partial_i\delta\rho_r+4\rho_r\partial_i\Phi].
\end{equation}

Equation of motion for perturbation of the scalar field can be calculated by variation of (\ref{39.5}) as $\delta(\nabla_{\mu}T^{\mu\nu}_{\varphi})=-\delta Q^{\nu}$ giving
\begin{eqnarray}\label{42.2}
&&(1+{3H^2\over M^2})\ddot{\delta\varphi}+[(1+{3H^2\over M^2}+{2\dot{H}\over M^2})3H+\Gamma]\dot{\delta\varphi}
+\delta V'(\varphi)+\dot{\varphi}\delta\Gamma\\\nonumber
&&-(1+{3H^2\over M^2}+{2\dot{H}\over M^2}){\nabla^2\delta\varphi\over a^2}=\\\nonumber
&&-[2V'(\varphi)+3\Gamma\dot{\varphi}-{6H\dot{\varphi}\over M^2}(3H^2+2\dot{H})
-{6H^2\ddot{\varphi}\over M^2}]\Phi\\\nonumber
&&+(1+{9H^2\over M^2})\dot{\varphi}\dot\Phi+{2H\dot{\varphi}\over M^2}{\nabla^2\Phi\over a^2}
+3(1+{9H^2\over M^2}+{2\dot{H}\over M^2}+{2H\ddot{\varphi}\over M^2})\dot{\Psi}\nonumber \\
&&+{6H\dot{\varphi}\over M^2}\ddot{\Psi}-{2(\ddot{\varphi}+H\dot{\varphi})\over M^2}{\nabla^2\Psi\over a^2},
\end{eqnarray}
for the zeroth component. By using perturbation to the Einstein field equation $G_{\mu\nu}=-T_{\mu\nu}$ (note that we have taken $8\pi G=1$), one can obtain the evolution equation for perturbation parameters, which for the $0-0$ component is
\begin{eqnarray}\label{43}
&&-3H\dot{\Psi}-3H^2\Phi+{\nabla^2\Psi\over a^2}={1\over 2}\big[-(1+{18H^2\over M^2}){\dot{\varphi}}^2\Phi-{9H{\dot{\varphi}}^2\over M^2}\dot{\Psi} \\\nonumber
&&+{{\dot{\varphi}}^2\over M^2}{\nabla^2\Psi\over a^2}+\acute{V(\varphi)}\delta\varphi+(1+{9H^2\over M^2})\dot\varphi\dot{\delta\varphi}
-{2H\dot{\varphi}\over M^2}{\nabla^2{(\delta\varphi)}\over a^2}+\delta\rho_r\big],
\end{eqnarray}
and the $i-i$ components are
\begin{eqnarray}\label{44}
&&(3H^2+2\dot{H})\Phi+H(3\dot{\Psi}+\dot{\Phi})+{\nabla^2(\Phi-\Psi)\over 3a^2}+\ddot{\Psi}=\\\nonumber
&&{1\over 2}[({(3H^2+2\dot{H}){2\dot{\varphi}^2\over M^2}-{\dot{\varphi}}^2+{8H\dot{\varphi}\ddot{\varphi}\over M^2}})\Phi+
{3H{\dot{\varphi}}^2\over M^2}\dot{\Phi} \\\nonumber
&&+{{\dot{\varphi}}^2\over M^2}{\nabla^2\Phi\over 3a^2}+({3H{\dot{\varphi}}^2\over M^2}+{2\dot{\varphi}\ddot{\varphi}\over M^2})\dot{\Psi}+{{\dot{\varphi}}^2\over M^2}\ddot{\Psi}+{{\dot{\varphi}}^2\over M^2}{\nabla^2\Psi\over 3a^2} \\\nonumber
&&-\acute{V(\varphi)}\delta\varphi-[(-1+{3H^2\over M^2}+{2\dot{H}\over M^2})\dot{\varphi}+{2H\ddot\varphi\over M^2}]\dot{\delta\varphi}\\\nonumber
&&-{2H\dot{\varphi}\over M^2}\ddot{\delta\varphi}
+{2(\ddot{\varphi}+H\dot{\varphi})\over M^2}{\nabla^2{(\delta\varphi)}\over 3a^2}+\delta P_r].
\end{eqnarray}
By relation $-H\partial_i\Phi-\partial_i\dot{\Psi}={1\over 2}(\rho+P)\partial_i\delta u $, from $0-i$ component of field equation we have
\begin{eqnarray}\label{45}
&&H\Phi+\dot{\Psi}={1\over 2}[{3H\dot{\varphi}^2\over M^2}\Phi+{\dot{\varphi}^2\over M^2}\dot{\Psi}+(1+{3H^2\over M^2})\dot{\varphi}\delta\varphi-{2H\dot{\varphi}\over M^2}\dot{\delta\varphi}\\\nonumber
&&+(\rho_r+P_r)\delta u].
\end{eqnarray}
These six equations (\ref{42}-\ref{45}), generally describe the evolution of perturbations.

We consider the quantities in momentum space via Fourier transform,  therefore the spatial parts of these quantities are $e^{ikx}$ where $k$ is the wave number of the corresponding mode. So by replacing $\partial_j\rightarrow ik_j $ and $\nabla^2\rightarrow-k^2 $, and defining
\begin{equation}\label{47}
\delta u={-a\over k}v e^{ikx},
\end{equation}
we can write the equation (\ref{42.1}) as
\begin{equation}\label{48}
\rho_r\dot{v}+\dot{\rho_r}v+4H\rho_r v={k\over a}[\rho_r\Phi+{\delta\rho_r\over4}+{3\over4}\Gamma\dot{\varphi}\delta\varphi].
\end{equation}
During warm inflation the background and perturbation satisfy the slow roll approximation. In other words the background and perturbations vary slowly in time (e.g. $\dot{\Phi}\ll H\Phi$). We consider modes with wavenumbers satisfying  ${k\over a}\ll H$. By applying this conditions to the equation (\ref{42}) and considering the high friction regime, we obtain
\begin{equation}\label{49}
{\delta\rho_r\over \rho_r} =-\Phi+{\delta\Gamma\over \Gamma}.
\end{equation}
Similarly, (\ref{48}) reduces to
\begin{equation}\label{50}
v={k\over 4aH}[\Phi+{\delta\rho_r\over4\rho_r}+{3\Gamma\dot{\varphi}\delta\varphi\over4\rho_r}],
\end{equation}
and the equation (\ref{42.2}) takes the form
\begin{eqnarray}\label{51}
&&[(1+{3H^2\over M^2})3H+\Gamma]\dot{\delta\varphi}
+\delta V'(\varphi)+\dot{\varphi}\delta\Gamma=\\\nonumber
&&-[2V'(\varphi)+3\Gamma\dot{\varphi}-{3H^2\over M^2}(6H\dot{\varphi})]\Phi.
\end{eqnarray}
We derive also
\begin{eqnarray}\label{52}
H\Phi={1\over 2}[{3H\dot{\varphi}^2\over M^2}\Phi+(1+{3H^2\over M^2})\dot{\varphi}\delta\varphi-{4a\over 3k}\rho_rv].
\end{eqnarray}
From relations (\ref{49}-\ref{52}) we can calculate $\delta\varphi$ as a function of $H$, $\Gamma$, and $V(\varphi)$,
\begin{equation}\label{53}
\delta\varphi\approx C V' \exp{(\Im(\varphi))},
\end{equation}
where $\Im(\varphi)$ is defined as
\begin{equation}\label{54}
\Im(\varphi)\equiv-\int{\Big({\Gamma'\over\Gamma}{r\over 1+r}+{V'\over V}{2+5r\over 2(r+1)^2}
{\Big[1+{3r\over4}-{\beta r\over16(1+r)}\Big]}\Big)}d\varphi.
\end{equation}

The density perturbation is then \cite{Herrera2},\cite{Herrera},\cite{Berera2}
\begin{equation}\label{56}
\delta_H={16\pi\over 5}{\exp{(-\Im(\varphi))}\over V'}\delta\varphi.
\end{equation}
In this relation $\delta\varphi$ is the fluctuation of the scalar field during the warm inflation \cite{Berera}, \cite{Herrera}
\begin{equation}\label{57}
\delta\varphi^2={k_FT\over2\pi^2},
\end{equation}
where $k_F$ is the freeze out scale. To calculate $k_F$, we must determine the time at which the damping rate of relation (\ref{42.2}) falls below the expansion rate $H$. At the freeze out time, $t_F$, the freeze out wavenumber, $k_F={k\over a(t_F)}$, is given by
\begin{equation}\label{58}
k_F=\sqrt{\Gamma H+3H^2(1+{3H^2\over M^2})}=\sqrt{3H^2U(1+r)},
\end{equation}
therefore the density perturbation becomes
\begin{equation}\label{59}
\delta_H^2= \left({128\over 25}\right)\left({\exp(-2\Im(\varphi))\over{V'(\varphi)}^2}\right)\sqrt{3H^2U(1+r)}T.
\end{equation}
The spectral index for the scalar perturbation is given by
 \begin{equation}\label{60}
n_s-1={d\ln{\delta_H^2}\over d\ln{k}},
\end{equation}
where this derivative is computed at the horizon crossing $k\approx aH$. Finally we obtain
\begin{eqnarray}\label{61}
&&n_s-1={2\eta\over(1+r)}-{\delta\over2(1+r)}-{\beta(1+5r)\over2{(1+r)}^2}-{\delta(2+5r)(4+3r)\over2(1+r)^2}\\\nonumber
&&+{\delta\beta r(2+5r)\over8(1+r)^4}
\end{eqnarray}

Using (\ref{23}) and (\ref{24}), one can see that the slow roll parameters are
\begin{equation}\label{62}
\delta\sim {n^2\over 2}\alpha,\qquad  \eta\sim n(n-1)\alpha,\qquad  \beta\sim pn\alpha,
\end{equation}
where
\begin{equation}\label{63}
\alpha={M^2\over \lambda}\varphi^{-(n+2)}.
\end{equation}
For $r\gg1$, we have
\begin{equation}\label{65}
\delta_H^2=\left({128\over 25\times 3^{1\over 4}}\right)\left({V^4\Gamma^{5\over 2}\over {V'}^2}\right)T.
\end{equation}
With our power law choices for the potential and dissipation coefficient, (\ref{65}) reduces to
\begin{equation}\label{66}
\delta_H^2= \left({128\lambda^2\over 25\times3^{1\over4}n^2}\right){\left({\Gamma_0\over\varphi_0^p}\right)}^{5\over2}\varphi^{(2n+2+{5p\over2})}T.
\end{equation}

The spectral index is
\begin{equation}\label{67}
n_s-1={2\eta\over r}-{8\delta\over r}-{5\beta\over 2r}.
\end{equation}
We can rewrite this relation as
\begin{equation}\label{68}
n_s-1=-{n\alpha\over r}\left(2n+2+{5p\over 2}\right),
\end{equation}
where $r$  is given by
\begin{equation}\label{69}
r={\Gamma_0 M^2\over\sqrt{3}\varphi_0^p\lambda^{3\over2}}\varphi^{(p-{3n\over2})}.
\end{equation}

By inserting the value of  $\varphi$ at the horizon crossing in (\ref{68}) we get
\begin{equation}\label{69.1}
n_s-1=-{n\sqrt{3\lambda}\varphi_0^p\over \Gamma_0}\left(2n+2+{5p\over 2}\right)\varphi^{-(p+2-{n\over2})}.
\end{equation}

\section{Evolution of the universe and temperature of the warm inflation}

In this section, using our previous results,  we intend to calculate the temperature of warm inflation as a function of observational parameters via the method introduced in \cite{mir}.
By the temperature of warm inflation, we mean the temperature of the universe at the end of warm inflation.
For this purpose, we divide the evolution of the universe into three parts as follows\\
$I-$ from $t_\star$ (horizon exit) until the end of slow roll warm inflation, denoted by $t_e$. In this era, the potential of the scalar field is the dominant term in the energy density.\\
$II-$ from $t_e$ until recombination era, denoted by $t_{rec}$.\\
$III-$ from $t_{rec}$ until the present time $t_0$.\\
Therefore the number of e-folds from horizon crossing until now becomes
\begin{eqnarray}\label{69.2}
\mathcal{N}&=&\ln{({a_0\over a_\star})}=\ln{({a_0\over a_{rec}})}+\ln{({a_{rec}\over a_{e}})}+\ln{({a_e\over a_{\star}})} \nonumber\\
&=&\mathcal{N}_{I}+\mathcal{N}_{II}+\mathcal{N}_{III}
\end{eqnarray}

\subsection{Slow roll}

During the slow roll warm inflation, the scalar field rolls down to the minimum of the potential and ultra relativistic particles are generated . In this period the positive potential energy of the scalar field is dominant and  therefore expansion of the universe is accelerated.
By relations (\ref{34.1}) and (\ref{69.1}), for high damping term $r\gg1$, the number of e-folds during warm inflation becomes
\begin{equation}\label{69.3}
\mathcal{N}_I={2n+2+{5p\over2}\over(p+2-{n\over2})(1-n_s)}.
\end{equation}
We need to calculate scalar field and temperature at the end of slow roll.
Inflation ends at the time when $r(\varphi_{end})\sim\delta(\varphi_{end})$. From equations (\ref{62}) and (\ref{69}), we can calculate the scalar field at the end of inflation as
\begin{equation}\label{69.4}
\varphi_{end}^{-(p+2-{n\over2})}\simeq {2\Gamma_0\over n^2\sqrt{3\lambda}\varphi_0^p}.
\end{equation}

At the end of inflation the radiation energy density gains the same order as the energy density of the scalar field
\begin{equation}\label{69.5}
\rho_{end}\simeq  V(\varphi_{end})=\lambda{\Big({n^2\sqrt{3\lambda}\varphi_0^p\over2\Gamma_0}\Big)}^{n\over p+2-{n\over2}}.
\end{equation}
From equation (\ref{28}) we deduce that the temperature of the universe at the end of inflation is
\begin{equation}\label{69.6}
T_{end}\simeq \left({30\lambda\over g \pi^2}\right)^{1\over4}\left({2\Gamma_0\over n^2\sqrt{3\lambda}\varphi_0^p}\right)^{-n\over4(p+2-{n\over2})}.
\end{equation}

\subsection{Radiation dominated and recombination eras}

At the end of the warm inflation, the universe enters a radiation dominated epoch. During this era the universe is filled of ultra-relativistic particles which are in thermal equilibrium, and experiences an adiabatic expansion during which the entropy per comoving volume is conserved: $dS=0$ \cite{kolb}. In this era the entropy
density,  $s=Sa^{-3}$, is derived as \cite{kolb}
\begin{equation}\label{70}
s={2\pi^2\over 45}g T^3,
\end{equation}
So we have
\begin{equation}\label{71}
{a_{rec}\over a_{end}}={T_{end}\over T_{rec}}\left({g_{end}\over g_{rec}}\right)^{1\over 3}.
\end{equation}

In the recombination era, $g_{rec}$ is related to photons degrees of freedom and as a consequence  $g_{rec}=2$. Hence
\begin{equation}\label{72}
\mathcal{N}_{II}= \ln\left({T_{end}\over
T_{rec}}\left({g_{end}\over 2}\right)^{1\over 3}\right).
\end{equation}

By the expansion of the universe, the temperature diminishes:
$T(z)=T(z=0)(1+z)$, where $z$ is the redshift parameter. So we can state $T_{rec}$ in terms of $T_{CMB}$ as
\begin{equation}\label{73}
T_{rec}=(1+z_{rec})T_{CMB}.
\end{equation}
We have also
\begin{equation}\label{74}
{a_0\over a_{rec}}=(1+z_{rec}),
\end{equation}
hence
\begin{equation}\label{75}
\mathcal{N}_{II}+\mathcal{N}_{III}=\ln\left({T_{end}\over
T_{CMB}}\left({g_{end}\over 2}\right)^{1\over3}\right).
\end{equation}

\subsection{Temperature in the warm inflation}

We have determined the number of e-folds appearing in the the right
hand side of (\ref{69.2}). To determine the warm inflation temperature we require
to determine  $\mathcal{N}$ in (\ref{69.2}). By assuming $a_0=1$, the number of
e-folds from the horizon crossing until the present time is
obtained as $\mathcal{N}=\ln(\Delta)$, where

\begin{equation}\label{76}
\Delta={1\over a_*}={H_*\over k_0}={V(\varphi_\star)^{1\over2}\over\sqrt{3}k_0}.
\end{equation}
By relations (\ref{69.2},\ref{69.3},\ref{75}) we can  obtain $T_{end}$ as
\begin{equation}\label{76.1}
T_{end}=T_{CMB}{\left({2\over g_{end}}\right)}^{1\over3}\exp{\left(\mathcal{N}-{(2n+2+{5p\over2})\over(p+2-{n\over2})(1-n_s)}\right)},
\end{equation}
which by using  relation (\ref{76}) can be expressed as
\begin{equation}\label{78}
T_{end}=T_{CMB}{\left({2\over g_{end}}\right)}^{1\over3}{\lambda^{1\over2}\varphi_\star^{n\over2}\over\sqrt{3}k_0}\exp{\left(-{(2n+2+{5p\over2})\over(p+2-{n\over2})(1-n_s)}\right)}.
\end{equation}

With the help of the relations $\mathcal{P}_s(k_0)={25\over 4}\delta^2_H(k_0)$ ($k_0$ is the pivot scale) and (\ref{66}) we express the power spectrum as
\begin{equation}\label{77}
\mathcal{P}_s(k_0)\approx {\left({32\over 3^{{1\over4}}n^2}\right)}
{\left({\Gamma_0\over\varphi_0^p}\right)}^{5\over2}\lambda^2\varphi_\star^{(2n+2+{5p\over2})}T_\star.
\end{equation}
In the above equation $T_\star$ is the temperature of the universe at the horizon crossing where the relation (\ref{29}) holds, thus
\begin{equation}\label{77.1}
\mathcal{P}_s(k_0)\approx {32\lambda^{19\over8}\over n^{3\over2}}{\left({\left({\Gamma_0\over\varphi_0^p}\right)}^9\times {15\over 2\sqrt{3} g_{end}\pi^2}\right)}^{1\over4}\varphi_\star^{({19n+12+18p\over8})}.
\end{equation}
From (\ref{69.1}) and (\ref{77.1}) we have
\begin{equation}\label{r21}
\varphi_*=\left({\mathcal{P}_s(k_0)\over \Omega (1-n_s)^{19\over 4}\left({\Gamma_0\over \varphi_0^p}\right)^7}\right)^{1\over 7p+1},
\end{equation}
where
\begin{equation}\label{r22}
\Omega={2^{9.5}\times 3^{-2.25}\times 5^{0.25}\over  n^{{25\over 4}}(4n+5p+4)^{{19\over 4}}(\pi^2g_{end})^{1\over 4}}.
\end{equation}
In addition from (\ref{69.1}) and (\ref{77.1}) and (\ref{r21})
\begin{equation}\label{r23}
\sqrt{\lambda}={\mathcal{P}_s(k_0)^{p-{n\over 2}+2\over 7p+1}\left({\Gamma_0\over \varphi_0^p}\right)^{7n-26\over 2(1+7p)}(1-n_s)^{18p+19n-68\over 8(7p+1)}\over \sqrt{3}n\Omega^{p+2-{n\over 2}\over 7p+1}(2n+{5p\over 2}+2)}.
\end{equation}

By inserting (\ref{r21}) and (\ref{r23}) in (\ref{78}), we derive
\begin{equation}\label{r24}
T_{end}=B{T_{CMB}\over k_0}(1-n_s)^{9p-34\over 28p+4}\mathcal{P}_s(k_0)^{p+2\over 7p+1}\left({\Gamma_0\over \varphi_0^p}\right)^{-13\over 7p+1}\exp\left(-{2n+{5p\over 2}+2\over (p-{n\over 2})(1-n_s)}\right),
\end{equation}
where $B$ is defined by
\begin{equation}\label{r25}
B={2^{1\over 3}\Omega^{-{p+2\over 7p+1}}\over 3ng_{end}^{1\over 3}(2n+{5p\over 2}+2)}.
\end{equation}

We use (\ref{69.6}) and (\ref{r23}) to obtain a second equation for the temperature in terms of dissipation factor as
\begin{equation}\label{r251}
T_{end}=C(1-n_s)^{-{(p+2)(18p+19n-68)\over 8(7p+1)(-2p+n-4)}}P_s(k_0)^{2+p\over 2(7p+1)}\left({\Gamma_0\over \varphi_0^p}\right)^{-{13\over 2(7p+1)}},
\end{equation}
where $C$ is defined by
\begin{equation}\label{r252}
C=\left({30\over \pi^2g_{end}}\right)^{1\over4}\left({2\over \sqrt{3}n^2}\right)^{-{n\over 4\left(p-{n\over 2}+2\right)}}\Omega^{-{{p+2}\over 2(7p+1)}}\left(\sqrt{3}n\left(2n+{5p\over 2}+2\right)\right)^{-{{p+2\over 2p-n+4}}}.
\end{equation}

By combining (\ref{r251}) and (\ref{r24}), we can determine the temperature in terms $p$, $n$, and the spectral index
\begin{equation}\label{r26}
T_{end}=K{k_0\over T_{CMB}}(1-n_s)^{n\over 2p+4-n}\exp\left({2n+{5p\over 2}+2\over (p-{n\over 2}+2)(1-n_s)}\right)M_P,
\end{equation}
where
\begin{equation}\label{r261}
K={\sqrt{90}(4n+5p+4)\over 2^{-{2(p+n+2)\over 3(n-2p-4)}}\pi g_{end}^{1\over 6}n^{n\over -2p+n-4}\left(2n+{5p\over 2}+2\right)^{2(p+2)\over 2p-n+4}}.
\end{equation}
$M_P=2.4\times 10^{18}GeV=8\pi G$ is the reduced Planck mass. Hereafter we reset the natural units.
(\ref{r26}) is completely different from the result obtained for temperature in reheating era in the ordinary (cold) minimal inflation obtained in \cite{mir} for quadratic potential
\begin{equation}\label{r27}
T_{end}=0.085\sqrt{{(1-n_s)\over \mathcal{P}_s}}\left(k_0\over T_{CMB}\right)^3\exp\left({6\over 1-n_s}\right)M_P.
\end{equation}

Up to first order Taylor expansion, the relative uncertainty in our result is
\begin{equation}\label{r28}
{\sigma(T_{end})\over T_{end}}=\sqrt{{\sigma^2(n_s)\over T_{end}^2}\left({\partial T\over \partial n_s}\right)^2}.
\end{equation}
The two conditions that we have used for calculation of the temperature, i.e.  $r\gg1$ and ${H^2\over M^2}\gg1$, lead to
\begin{equation}\label{r29}
{(2n+{5p\over 2}+2)^3\over 3n\Omega^{-2p-6\over 7p+1}}\mathcal{P}_s(k_0)^{-{2p-6\over 7p+1}}(1-n_s)^{51-23p\over 2(7p+1)}\left({\Gamma_0\over \varphi_0^p}\right)^{40\over 7p+1}M^2\gg M_P^{42-26p\over 7p+1},
\end{equation}
and
\begin{equation}\label{r30}
{g_{end}\pi^2\over 90}T_{end}^4\gg M^2M_P^2,
\end{equation}
respectively. (\ref{r29}) was derived from (\ref{69}) and (\ref{r30}) was obtained using ${1\over 3M_P^2}\rho_r\gg M^2$.

To calculate $T_{end}$, we set $g_{end}=106.75$, which is the ultra relativistic degrees of freedom at the electroweak energy scale. From PLANK 2013 for pivot scale $k_0=0.05Mpc^{-1}$ in one sigma level, we set $\mathcal{P}_s(k_0)=(2.20\pm0.056)\times10^{-9}$
and $n_s=0.9608\pm0.0054$ \cite{plank}. Note that ${k_0\over T_{CMB}}={0.05 Mpc^{-1}\over 2.725 K}=0.05\times 10^{-26}$. After fixing these parameters, the temperature depends entirely on $p$ and $n$. As an example if one takes  $p=-1$ (this choice gives a positive power for $\Gamma_0$ in (\ref{r24})) and  $n=0.8$ (for non integer values of $n$ see (\cite{end})), he obtains
\begin{equation}
5.01\times 10^7GeV <T_{end} < 2.11\times 10^{13}GeV.
\end{equation}
For $n_s=0.9608$,  the temperature is $T_{end}=1.32\times 10^{10} GeV$ whose relative uncertainty is ${\sigma(T_{end})\over T_{end}}=6.35$.

The range of the temperature, must lie below the upper bound scale assumed in the literature
which is about the GUT scale $T_{max.}\simeq 10^{16}GeV$. By considering the big bang nucleosynthesis (BBN), and on the base of the data derived
from large scale structure and also cosmic microwave background (CMB), a lower bound  $T_{min.}\simeq 4MeV$, is obtained in \cite{han} which is consistent with our example.

\section{Summary}
We considered warm inflation in the framework of non minimal derivative coupling model in high friction regime. After an introduction to the model, we studied the slow roll conditions and e-folds number and then specified them in terms of the parameters of the model for a power law potential and a general power law dissipation factor.  By studying The cosmological perturbations, we obtained the power spectrum and the spectral index. We used these quantities to determine the temperature of the universe in terms of $T_{CMB}$ and the spectral index.

\end{document}